\documentclass[11pt]{article}
\newcommand{\nc}{\newcommand}
\nc{\bib}{\bibitem}
\nc{\al}{\alpha}
\nc{\g}{\gamma}
\nc{\G}{\Gamma}
\nc{\D}{\Delta}
\nc{\eps}{\epsilon}
\nc{\la}{\lambda}
\nc{\La}{\Lambda}
\nc{\var}{\varphi}
\nc{\pa}{\partial}
\nc{\nn}{\nonumber \\ }
\nc{\hf}{\frac{1}{2}}         
\nc{\dz}{\frac{dz}{2\pi i}}
\nc{\bin}[2]{\left (\begin{array}{c} {#1}\\ {#2} \end{array}\right )}
\nc{\ben}{\begin{equation}}
\nc{\een}{\end{equation}}
\nc{\bea}{\begin{eqnarray}}
\nc{\eea}{\end{eqnarray}}
\nc{\bra}[1]{\langle {#1}|}
\nc{\ket}[1]{|{#1}\rangle}
\newcommand{\Z}{\mbox{$Z\hspace{-2mm}Z$}}
\nc{\C}{\mbox{\hspace{1.24mm}\rule{0.2mm}{2.5mm}\hspace{-2.7mm} C}}
\nc{\Nat}{\mbox{\hspace{.04mm}\rule{0.2mm}{2.8mm}\hspace{-1.5mm} N}}

\nc{\NP}[1]{Nucl.\ Phys.\ {\bf #1}}
\nc{\PL}[1]{Phys.\ Lett.\ {\bf #1}}
\nc{\CMP}[1]{Commun.\ Math.\ Phys.\ {\bf #1}}
\nc{\PR}[1]{Phys.\ Rev.\ {\bf #1}}
\nc{\PRL}[1]{Phys.\ Rev.\ Lett.\ {\bf #1}}
\nc{\PTP}[1]{Prog.\ Theor.\ Phys.\ {\bf #1}}
\nc{\PTPS}[1]{Prog.\ Theor.\ Phys.\ Suppl.\ {\bf #1}}
\nc{\MPL}[1]{Mod.\ Phys.\ Lett.\ {\bf #1}}
\nc{\IJMP}[1]{Int.\ Jour.\ Mod.\ Phys.\ {\bf #1}}
\nc{\IM}[1]{Invent.\ Math.\ {\bf #1}}
\nc{\SJNP}[1]{Sov. J. Nucl. Phys.\ {\bf #1}}
\nc{\JHEP}[1]{J.\ High\ Energy Phys.\ {\bf #1}}

\def\tri#1#2#3#4#5#6#7#8#9{\matrix{\quad\cr #4\cr
	#3\quad#5\cr #2~\qquad #6\cr #1\quad #9\quad#8\quad#7\cr 
         \quad\cr}}

\def\vvdots{\mathinner{\mkern1mu\raise1pt\vbox{\kern7pt\hbox{.}}\mkern2mu
 \raise4pt\hbox{.}\mkern2mu\raise7pt\hbox{.}\mkern1mu}}

\def\max{{\rm max}}
\def\min{{\rm min}}

\begin{document}

\topmargin -5mm
\oddsidemargin 5mm

\begin{titlepage}
\setcounter{page}{0}

\vspace{8mm}
\begin{center}
{\huge Affine $su(3)$ and $su(4)$ fusion multiplicities}
\\[.4cm]
{\huge as polytope volumes}

\vspace{15mm}
{\large J{\o}rgen Rasmussen}\footnote{rasmussj@cs.uleth.ca; supported in part
by a PIMS Postdoctoral Fellowship and by NSERC} and 
{\large Mark A. Walton}\footnote{walton@uleth.ca; supported in part by NSERC}
\\[.2cm]
{\em Physics Department, University of Lethbridge,
Lethbridge, Alberta, Canada T1K 3M4}

\end{center}

\vspace{10mm}
\centerline{{\bf{Abstract}}}
\vskip.4cm
\noindent
Affine $su(3)$ and $su(4)$ fusion multiplicities are characterised 
as discretised volumes of certain convex polytopes. The volumes 
are measured explicitly, resulting in multiple sum formulas. 
These are the first polytope-volume formulas for higher-rank fusion
multiplicities. The associated
threshold levels are also discussed. For any simple Lie algebra we derive
an upper bound on the threshold levels using a refined version of the 
Gepner-Witten depth rule. 
\end{titlepage}
\newpage
\renewcommand{\thefootnote}{\arabic{footnote}}
\setcounter{footnote}{0}

\section{Introduction}

We are interested in describing fusions of 
integrable highest weight modules of affine Lie algebras.
Let $M_\la$ denote an integrable highest weight module of an untwisted
affine Lie algebra. The affine weight is uniquely specified by the highest 
weight $\la$ of the simple horizontal subalgebra (the underlying Lie algebra),
and the affine level $k$. 
Fusion of two such modules may be written as 
\ben
 M_\la\times M_\mu=\sum_\nu\ N_{\la,\mu}^{(k)\ \nu} M_\nu\ \ ,
\label{MM}
\een
where $N_{\la,\mu}^{(k)\ \nu}$ is the fusion multiplicity.
This is equivalent to studying the more 
symmetric problem of determining the multiplicity of the singlet 
in the expansion of the triple fusion
\ben
 M_\la\times M_\mu\times M_\nu\supset N_{\la,\mu,\nu}^{(k)}M_0\ .
\label{MMM}
\een
If $\nu^+$ denotes the weight conjugate to 
$\nu$, we have $N_{\lambda,\mu,\nu}^{(k)} = N_{\lambda,\mu}^{(k)\ \nu^+}$.  

The associated and level-independent tensor product multiplicity 
is denoted $T_{\la,\mu,\nu}$:
\ben
 M_\la\otimes M_\mu\otimes M_\nu\supset T_{\la,\mu,\nu}M_0\ .
\label{T}
\een
It is related to the fusion multiplicity as
\ben
 T_{\la,\mu,\nu}=\lim_{k\rightarrow\infty}N_{\la,\mu,\nu}^{(k)}\ .
\label{TNlim}
\een
It has been conjectured \cite{CMW} that the fusion multiplicities
are uniquely determined from the tensor product multiplicities and the
associated multi-set of minimum levels $\{t\}$ 
at which the various couplings first appear.
Therefore, to the triplet $(\la,\mu,\nu)$ there correspond
$T_{\la,\mu,\nu}$ distinct couplings, hence $T_{\la,\mu,\nu}$ values
of $t$, one for each distinct coupling. These values are called
threshold levels. The threshold levels associated to two different
couplings may be identical. That justifies the use of the notion multi-set 
of threshold levels to describe the general case. 
The number of different couplings with the
same threshold level $t$ is called the threshold multiplicity,
$n_{\la,\mu,\nu}^{(t)}$, and may be
expressed in terms of the fusion multiplicities:
\ben
 n_{\la,\mu,\nu}^{(t)}=N_{\la,\mu,\nu}^{(t)}-N_{\la,\mu,\nu}^{(t-1)}\ .
\label{tmult}
\een

In Ref. \cite{BZ} Berenstein and Zelevinsky showed that an $su(r+1)$ 
tensor product multiplicity is equal to the number of possible
triangular arrangements of non-negative integers subject to certain
constraints, referred to as BZ triangles. A triangle without the constraint
that all the integer entries should be {\em non-negative}, 
is called a generalised BZ triangle \cite{RW1}. 
The set of generalised triangles associated to a particular
product (\ref{T}) spans an $\hf r(r-1)$-dimensional lattice. Any triangle
${\cal T}$ in the lattice
may be expressed in terms of an initial one ${\cal T}_0$
plus an integer linear combination
of so-called (basis) virtual triangles ${\cal V}_l$:
\ben
 {\cal T}={\cal T}_0+\sum_{l=1}^{r(r-1)/2}v_l{\cal V}_l\ .
\een
Re-imposing the constraint that all entries must be non-negative, 
results in a 
set of inequalities in the coefficients $v_l$ defining a convex polytope.
Its discretised volume (i.e., the number of integer points enclosed by it)
is the tensor product multiplicity \cite{RW1}.

This idea was generalised to higher-point couplings in \cite{RW2},
and to affine $su(2)$ fusions in \cite{RW3} (that work also describes the
extension to higher-point fusions and higher genus).
The objective of the present work is the extension to affine $su(3)$ and
$su(4)$ fusions. Thus, the affine $su(3)$ and $su(4)$ fusion multiplicities 
are characterised 
as discretised volumes of certain polytopes. The volumes are subsequently
measured explicitly, resulting in multiple sum formulas for
the fusion multiplicities. 

We also discuss the associated threshold levels, and for $su(3)$ and $su(4)$
work out explicitly the
minimum threshold level $t^\min$, and the maximum threshold level $t^\max$.
In the case of $su(4)$ these are new results. By construction, $t^\max$ 
is the minimum level for which the (non-vanishing) tensor product
multiplicity and the fusion multiplicity coincide. It is therefore of 
particular interest to know $t^\max$. Using a refined depth rule
applicable to all simple Lie algebras, we find an upper bound on $t^\max$.
Motivated by our results for $su(4)$ (and the known results for $su(2)$
and $su(3)$), we conjecture that the upper bound on $t^\max$ is 
saturated. 

\section{$su(3)$ fusion multiplicities}

An $su(3)$ BZ triangle, describing a particular coupling 
(to the singlet $M_0$) associated to the
triple tensor product $M_\lambda\otimes M_\mu\otimes M_\nu$, 
is a triangular arrangement of 9 non-negative integers:
\ben
 \matrix{\quad\cr m_{13}\cr
	n_{12}~~\quad l_{23}\cr
 m_{23}~\quad\qquad ~~m_{12}\cr
 n_{13}~\quad l_{12} \qquad n_{23} \quad~ l_{13} \cr \quad\cr}
\label{trithree}
\een
These integers are related to the Dynkin labels of the three integrable
highest weights by 
\ben
\begin{array}{llll}
 &m_{13}+n_{12}=\lambda_1\ ,\ \ &n_{13}+l_{12}=\mu_1\ ,\ \
 &l_{13}+m_{12}=\nu_1\ ,\nn
 &m_{23}+n_{13}=\lambda_2\ ,\ \ &n_{23}+l_{13}=\mu_2\ ,\ \
 &l_{23}+m_{13}=\nu_2\ .
\end{array}
\label{outthree}
\een 
We call these relations outer constraints.
The entries further satisfy the so-called hexagon identities
\ben
\begin{array}{l}
 n_{12}+m_{23}=n_{23}+m_{12}\ ,\nn 
 m_{12}+l_{23}=m_{23}+l_{12}\ ,\nn 
 l_{12}+n_{23}=l_{23}+n_{12}
\end{array}
\label{hexthree}
\een
of which only two are independent.

In the case of $su(3)$ there is one basis virtual triangle
\ben
 {\cal V}=\tri{1}{\bar1}{\bar1}{1}{\bar1}{\bar1}{1}{\bar1}{\bar1}
\label{Vthree}
\een
where ${\bar1}\equiv-1$.
It is easy to work out an initial triangle (a choice of
initial triangle valid for all $su(r+1)$, may be found in \cite{RW1}):
\ben
 {\cal T}_0=\matrix{\quad\cr N_2'\cr
	n_2~~\quad N_2\cr
 \la_2~\quad\qquad ~~N_1'\cr
 0\qquad \mu_1\quad~ n_1 \quad~ N_1 \cr \quad\cr}
\label{T03}
\een
where
\bea
 &&n_1=\la^2+\mu^2-\nu^1,\ n_2=\la^1+\mu^1-\nu^2,\nn
 &&N_1=-n_1+\mu_2,\ N_2=n_1-n_2+\mu_1,\ N_i'=\nu_i-N_i,\ i=1,2\ .
\eea
The dual Dynkin labels of the $su(3)$ weight $\la$ are 
\ben
 \la^1=\frac{1}{3}(2\la_1+\la_2),\ \ \ \la^2=\frac{1}{3}(\la_1+2\la_2)\ .
\een
In general, ordinary and dual Dynkin labels are defined by
\ben
 \la=\sum_{i=1}^r\la_i\La^i=\sum_{i=1}^r\la^i\al_i^\vee\ ,
\label{dual}
\een
where $\{\La^i\}$ and $\{\al_i^\vee\}$ are the sets of fundamental weights and
simple co-roots, respectively. $r$ is the rank of the Lie algebra. It is
$r=N-1$ for $su(N)$. The set of simple roots is denoted $\{\al_i\}$.

Any generalised triangle may now be expressed as
\ben
 {\cal T}={\cal T}_0+v{\cal V}\ .
\een
Re-imposing that all entries of ${\cal T}$ must be {\em non-negative}
integers, defines a convex polytope (in this case a line segment)
characterised by the inequalities
\ben
 0\leq N_2'+v,\ n_2-v,\ \la_2-v,\ v,\ 
   \mu_1-v,\ n_1-v,\ N_1+v,\ N_1'-v,\ N_2-v\ .
\label{ten3}
\een
Its discretised volume is the tensor product multiplicity $T_{\la,\mu,\nu}$.
Note the implicit consistency conditions 
\ben
 S_i\equiv\la^i+\mu^i+\nu^i\in\Z_\geq,\ \ \ i=1,2\ ,
\label{S3}
\een 
which must be respected to have a non-vanishing multiplicity.

It was shown in Ref. \cite{KMSW}  that one can assign the threshold value
\ben
 t=\max\{\la_1+\la_2+l_{13},\ \mu_1+\mu_2+m_{13},\ \nu_1+\nu_2+n_{13}\}\ .
\label{k03}
\een
to any given $su(3)$ BZ triangle (\ref{trithree}). This means that 
the fusion multiplicity may be described by supplementing the tensor 
product conditions (\ref{ten3}) by the affine condition
\ben
 k\geq t\ .
\label{affine}
\een
Thus, the discretised volume of the convex polytope defined by the 
inequalities
\bea
 0&\leq&N_2'+v,\ n_2-v,\ \la_2-v,\ v,\ 
   \mu_1-v,\ n_1-v,\ N_1+v,\ N_1'-v,\ N_2-v,\nn
  &&k-\la_1-\la_2-N_1-v,\ k-\mu_1-\mu_2-N_2'-v,\ k-\nu_1-\nu_2-v\ ,
\label{cp3}
\eea
is the {\em fusion} multiplicity $N_{\la,\mu,\nu}^{(k)}$.
The volume is easily measured explicitly, and we are left with the 
sum:
\bea
 N_{\la,\mu,\nu}^{(k)}&=&\sum_{v=v^L}^{v^U}1\ ,\nn
 v^L&=&\max\{0,\ -\la^1+\la^2+\mu^1-\nu^2,\ 
   \la^2+\mu^1-\mu^2-\nu^1\}\ ,\nn
 v^U&=&\min\{\la_2,\ \mu_1,\ \la^2+\mu^2-\nu^1,\ 
   \la^1+\mu^1-\nu^2,\ \la^2+\mu^1-\mu^2+\nu^1-\nu^2,\nn 
 &&\ \ \ \ \ \ -\la^1+\la^2+\mu^1-\nu^1+\nu^2,\ k-\la^1+\mu^1-\mu^2-\nu^1,\nn 
 &&\ \ \ \ \ \ k-\la^1+\la^2-\mu^2-\nu^2,\ 
  k-\nu^1-\nu^2\}\ .
\label{sum3}
\eea
For ease of use, it is expressed entirely in terms of the level $k$ and
the dual and ordinary Dynkin labels.
This expression is of course not unique due to the {\em choice} of initial
triangle (\ref{T03}) and the many possible ways of re-writing it.
We recall the consistency conditions (\ref{S3}).

\subsection{Threshold levels}

As an application of (\ref{sum3})
we may address the question when the fusion multiplicity
is greater than any given (non-negative) integer $M$.
The answer is easily obtained since it corresponds to requiring that
$v^U-v^L\geq M$, and we find the 27 conditions
\bea
 M&\leq&\la_i,\ \mu_i,\ \nu_i,\ \ \ i=1,2,3\ ,\nn
 M&\leq&S_i-(\la_1+\la_2),\ S_i-(\mu_1+\mu_2),\ S_i-(\nu_1+\nu_2),\nn
   &&S_i-(\la_i+\mu_i),\ S_i-(\mu_i+\nu_i),\ S_i-(\nu_i+\la_i),\ \ \
     i=1,2,3\ ,\nn
 M&\leq&k-(\la_1+\la_2),\ k-(\mu_1+\mu_2),\ k-(\nu_1+\nu_2)\ ,\nn
 M&\leq&k-S_i+\la_i,\ k-S_i+\mu_i,\ k-S_i+\nu_i,\ \ \ i=1,2,3\ .
\label{nonv3}
\eea
This is particularly interesting when $M=0$.
It also allows us to re-express the fusion multiplicity itself.
{}From (\ref{nonv3}) it follows immediately that necessary
conditions for $N_{\la,\mu,\nu}^{(k)}>0$ are
$k\geq \la_1+\la_2,\mu_1+\mu_2,\nu_1+\nu_2,
 S_i-\min\{\la_i,\mu_i,\nu_i\}$, $i=1,2$, while the fusion multiplicity
is equal to the tensor product multiplicity (i.e., $N_{\la,\mu,\nu}^{(k)}$
is independent of the level $k$) when $k\geq \min\{S_1,S_2\}$.
Furthermore, according to (\ref{k03}) and the structure of the 
virtual triangle (\ref{Vthree}), adding a virtual triangle to a 
triangle increases the threshold level by one.
We conclude that the fusion multiplicity is
\bea
 N_{\la,\mu,\nu}^{(k)}&=&\left\{\begin{array}{lll}
   0&{\rm if}&k<t^\min\ {\rm or}\ t^\max<t^\min\\
   k-t^\min+1&{\rm if}&t^\min\leq k\leq t^\max\\
   T_{\la,\mu,\nu}=t^\max-t^\min+1\ \ &{\rm if}&t^\min\leq t^\max
   <k\end{array}\right.\nn
 t^\min&=&\max\{\la_1+\la_2,\ \mu_1+\mu_2,\ \nu_1+\nu_2,\ 
 S_1-\min\{\la_1,\ \mu_1,\ \nu_1\},\ S_2-\min\{\la_2,\ \mu_2,\ \nu_2\}\}\ ,\nn
 t^\max&=&\min\{S_1,\ S_2\}\ ,
\label{3bmw}
\eea
still provided (\ref{S3}). The set of threshold levels for the 
$T_{\la,\mu,\nu}$ distinct couplings is easily read off:
\ben
 \{t^\min,\ t^\min+1,...,t^\max\}\ .
\label{string}
\een
This confirms the result of \cite{BMW} where the expression (\ref{3bmw})
and the threshold level string (\ref{string}) first appeared. 

\section{$su(4)$ fusion multiplicities}

For $su(4)$ a BZ triangle is defined in terms of 18 non-negative integers:
\ben
 \matrix{m_{14}\cr
	n_{12}~~\quad l_{34}\cr
 m_{24}~\qquad\qquad ~~m_{13}\cr
 n_{13}\qquad l_{23}\qquad n_{23} \qquad l_{24}\cr
 m_{34}\qquad\qquad\quad m_{23}\qquad\qquad\quad m_{12} \cr
 n_{14}\qquad l_{12}\qquad n_{24}\ \ \quad l_{13}\quad~~~ n_{34}\qquad
 l_{14} \cr  }
\label{trifour}
\een
related to the Dynkin labels by
\ben
 \begin{array}{llll}
 &m_{14}+n_{12}=\lambda_1\ ,\ \ &n_{14}+l_{12}=\mu_1\ ,\ \
 &l_{14}+m_{12}=\nu_1\ ,\nn
 &m_{24}+n_{13}=\lambda_2\ ,\ \ &n_{24}+l_{13}=\mu_2\ ,\ \
 &l_{24}+m_{13}=\nu_2\ ,\nn
 &m_{34}+n_{14}=\lambda_3\ ,\ \ &n_{34}+l_{14}=\mu_3\ ,\ \
 &l_{34}+m_{14}=\nu_3\ .
\end{array}
\label{outfour}
\een
The $su(4)$ BZ triangle contains three hexagons with the associated 
constraints:
\ben
 \begin{array}{llll}
 &n_{12}+m_{24} =m_{13}+n_{23}\ ,\ \  
 & n_{13}+l_{23} =l_{12}+n_{24}\ ,\ \
 & l_{24}+n_{23} =l_{13}+n_{34}\ ,\nn
 &n_{12}+l_{34}=l_{23}+n_{23}\ ,\ \
 & n_{13}+m_{34} =n_{24}+m_{23}\ ,\ \
 & n_{23}+m_{23} =m_{12}+n_{34}\ ,\nn
 &m_{24}+l_{23} =l_{34}+m_{13}\ ,\ \
 & m_{34}+l_{12} = l_{23}+m_{23}\ ,\ \
 &l_{13}+m_{23} =l_{24}+m_{12}\ . 
\end{array}
\label{hexfour}
\een
Only 6 of these 9 hexagon identities are independent.

In the case of $su(4)$ the three basis virtual triangles 
${\cal V}_1$, ${\cal V}_2$ and ${\cal V}_3$ are
\bea
 {\cal V}_1=\matrix{1\cr
	\bar1~~\quad\bar1\cr
 \bar1~\quad\quad ~~\bar1\cr
 1~\quad\bar1\quad\bar1\quad ~1\cr
 0\qquad\quad 1\qquad\quad 0 \cr
 0~\quad 0~\quad 0~\quad 0~\quad 0~\quad 0 \cr}
\hspace{1cm} {\cal V}_2=\matrix{0\cr
	0~~\quad 0\cr
 1~\quad\quad ~~0\cr
 \bar1~\quad \bar1\quad 1 \quad ~0\cr
 \bar1\qquad\quad \bar1\qquad\quad 0 \cr
 1~\quad\bar1~\quad\bar1~\quad 1~\quad 0~\quad 0 \cr}\nn
 {\cal V}_3=\matrix{0\cr
	0~~\quad 0\cr
 0~\quad\quad ~~1\cr
 0~\quad 1\quad\bar1 \quad ~\bar1\cr
 0\qquad\quad\bar1\qquad\quad\bar1 \cr
 0~\quad 0~\quad 1~\quad\bar1~\quad\bar1~\quad 1\cr}\hspace{3cm}\mbox{}
\label{Vfour}
\eea
We make the following choice of initial triangle \cite{RW1}:
\ben
 {\cal T}_0= \matrix{N_3'\cr
	n_3~~\quad N_3\cr
 \la_2~\qquad\qquad ~~N_2'\cr
 0\qquad \mu_1\qquad n_2 \qquad N_2\cr
 \la_3\quad\qquad\quad~~ \la_3\quad\qquad\quad~~ N_1' \cr
 0\qquad \mu_1\qquad 0\ \ \quad~ \mu_2\qquad n_1\quad~~ N_1 \cr  }
\label{T04}
\een
where
\bea
 &&n_1=\la^3+\mu^3-\nu^1,\ n_2=\la^2+\mu^2-\nu^2,\ n_3=\la^1+\mu^1-\nu^3,\nn
 &&N_1=-n_1+\mu_3,\ N_2=n_1-n_2+\mu_2,\ N_3=n_2-n_3+\mu_1,\ N_i'=\nu_i-N_i,\ 
  i=1,2,3\ .\ \ \ {}
\eea
The dual Dynkin labels (\ref{dual}) of the $su(4)$ weight $\la$ are  
\ben
 \la^1=\frac{1}{4}(3\la_1+2\la_2+\la_3),\ \ \ 
 \la^2=\frac{1}{4}(2\la_1+4\la_2+2\la_3),\ \ \ 
 \la^3=\frac{1}{4}(\la_1+2\la_2+3\la_3)\ .
\een

Now, any generalised triangle may be written
\ben
 {\cal T}={\cal T}_0+\sum_{l=1}^3v_l{\cal V}_l\ ,
\label{T4}
\een
and the tensor product multiplicity $T_{\la,\mu,\nu}$ is the discretised 
volume of the (in general) three-dimensional convex polytope
\bea
 0&\leq&v_2,\ \mu_1-v_2,\ \la_3-v_2,\ -v_2+v_3,\ v_1-v_2,\ 
   \mu_2+v_2-v_3,\ \la_2-v_1+v_2,\nn
 &&\la_3+v_1-v_2-v_3,\ \mu_1-v_1-v_2+v_3,\ 
   n_1-v_3,\ n_2-v_1+v_2-v_3,\ n_3-v_1,\nn
 &&N_1+v_3,\ N_1'-v_3,\ N_2+v_1-v_3,\ N_2'-v_1+v_3,\ N_3-v_1,\ N_3'+v_1\ .
\label{ten4}
\eea
We have the consistency conditions 
\ben
 S_i\equiv\la^i+\mu^i+\nu^i\in\Z_\geq,\ \ \ i=1,2,3\ .
\label{S4}
\een

It was shown in Ref. \cite{BKMW} (and confirmed in \cite{BCM}) 
that one can assign the threshold level
\bea
 t&=&\max\{\la_1+\la_2+\la_3+l_{14},\ \mu_1+\mu_2+\mu_3+m_{14},\
  \nu_1+\nu_2+\nu_3+n_{14},\nn
 &&\ \ \ \ \ \ \ \la_1+\la_2+l_{14}+l_{24}+n_{14},\ 
   \la_2+\la_3+l_{14}+l_{13}+m_{14},\nn 
 &&\ \ \ \ \ \ \   \mu_1+\mu_2+m_{14}+m_{24}+l_{14},\ 
   \mu_2+\mu_3+m_{14}+m_{13}+n_{14},\nn  
 &&\ \ \ \ \ \ \   \nu_1+\nu_2+n_{14}+n_{24}+m_{14},\  
  \nu_2+\nu_3+n_{14}+n_{13}+l_{14},\nn
 &&\ \ \ \ \ \ \ l_{14}+m_{14}+n_{14}+[\hf(\la_2+\mu_2+\nu_2+l_{23}
    +m_{23}+n_{23}+1)]\}
\label{k04}
\eea
to any given $su(4)$ BZ triangle (\ref{trifour}). $[x]$ denotes the
integer value of $x$, i.e., the greatest integer less than or equal to $x$. 
This means that the fusion multiplicity may be described by supplementing 
the tensor product conditions (\ref{ten4}) by the affine condition
(\ref{affine}), with (\ref{k04}) as the threshold level $t$.

After a simple re-writing, the condition $k\geq t$ on the last line
of (\ref{k04}) becomes
\ben
 k\geq [\hf(\la^2+\mu^2+\nu^2+l_{14}+m_{14}+n_{14}+1)]\ .
\label{last}
\een
According to (\ref{T4}),
it involves the integer value of a possibly half-integer number depending
on the parameters $v_l$ defining the polytope. Thus, the polytope is in 
general not convex, and measuring its volume is not straightforward.
However, we observe that for integer $k$ and $B$, the condition
$k\geq [B/2]$ is equivalent to $k\geq (B-1)/2$. The condition (\ref{last})
may therefore be simplified as
\ben
 v_1+v_2+v_3\leq 2k-\la^1+\la^3+\mu^1-\mu^3-\nu^1-\nu^2-\nu^3\ .
\een
In conclusion, 
the discretised volume of the {\em convex} polytope defined by $v_1$,
$v_2$ and $v_3$ subject to the inequalities
\bea
 0&\leq&v_2,\ \mu_1-v_2,\ \la_3-v_2,\ -v_2+v_3,\ v_1-v_2,\ 
   \mu_2+v_2-v_3,\ \la_2-v_1+v_2,\nn
 &&\la_3+v_1-v_2-v_3,\ \mu_1-v_1-v_2+v_3,\ 
   n_1-v_3,\ n_2-v_1+v_2-v_3,\ n_3-v_1,\nn
 &&N_1+v_3,\ N_1'-v_3,\ N_2+v_1-v_3,\ N_2'-v_1+v_3,\ N_3-v_1,\ N_3'+v_1,\nn
 &&k-\la_1-\la_2-\la_3-N_1-v_3,\ k-\mu_1-\mu_2-\mu_3-N_3'-v_1,\ 
     k-\nu_1-\nu_2-\nu_3-v_2,\nn
 &&k-\la_1-\la_2-N_1-N_2-v_1-v_2,\ k-\mu_2-\mu_3-N_2'-N_3'-v_2-v_3,\nn
 &&k+\la_3-\nu_1-\nu_2-\nu_3-v_2-v_3,\ 
   k+\mu_1-\nu_1-\nu_2-\nu_3-v_1-v_2,\nn
 &&k-\nu_1-\nu_2-N_3'-v_1-v_3,\ k-\nu_2-\nu_3-N_1-v_1-v_3,\nn
 &&2k-\la^1+\la^3+\mu^1-\mu^3-\nu^1-\nu^2-\nu^3-v_1-v_2-v_3\ ,
\label{cp4}
\eea
is the fusion multiplicity $N_{\la,\mu,\nu}^{(k)}$.
The volume is easily measured explicitly, and we are left with the 
multiple sum:
\bea
 N_{\la,\mu,\nu}^{(k)}&=&\sum_{v_2=v_2^L}^{v_2^U}\ \sum_{v_1=v_1^L}^{v_1^U}\ 
    \sum_{v_3=v_3^L}^{v_3^U}\ 1,\nn
 v_3^L&=&\max\{v_2,\ -\mu_1+v_1+v_2,\ \la^3+\mu^2-\mu^3-\nu^1,\ 
    -\la^2+\la^3-\mu^1+\mu^2-\nu^2+\nu^3+v_1\},\nn
 v_3^U&=&\min\{\mu_2+v_2,\ \la_3+v_1-v_2,\ \la^3+\mu^3-\nu^1,\ 
     \la^2+\mu^2-\nu^2-v_1+v_2,\nn 
   &&\ \ \ \ \ \ \la^3+\mu^2-\mu^3+\nu^1-\nu^2,\ 
    -\la^2+\la^3-\mu^1+\mu^2-\nu^1+\nu^2+v_1,\nn 
   &&\ \ \ \ \ \ k-\la^1+\la^3+\mu^1-\mu^3-\nu^2-v_2,\ 
    k+\la_3-\nu_1-\nu_2-\nu_3-v_2,\nn 
   &&\ \ \ \ \ \ k-\la^1+\la^2+\mu^1-\nu^1-\nu^2-v_1,\ 
    k+\la^3+\mu^2-\mu^3-\nu^2-\nu^3-v_1,\nn 
   &&\ \ \ \ \ \ k-\la^1+\mu^2-\mu^3-\nu^1,\ 
     2k-\la^1+\la^3+\mu^1-\mu^3-\nu^1-\nu^2-\nu^3-v_1-v_2\},\nn
 v_1^L&=&\max\{v_2,\ -\la^1+\la^2+\mu^1-\nu^3\},\nn
 v_1^U&=&\min\{\la_2+v_2,\ \la^1+\mu^1-\nu^3,\ 
    -\la^1+\la^2+\mu^1-\nu^2+\nu^3,\ k-\la^1+\la^2-\mu^3-\nu^3,\nn
   &&\ \ \ \ \ \ k-\la^1+\la^3+\mu^1-\mu^3-\nu^2-v_2,\ 
  k+\mu_1-\nu_1-\nu_2-\nu_3-v_2\},\nn
 v_2^L&=&0,\nn
 v_2^U&=&\min\{\mu_1,\ \la_3,\ k-\nu_1-\nu_2-\nu_3\}\ .
\label{sum4}
\eea
For ease of use, it is expressed entirely in terms of the level $k$ and
the dual and ordinary Dynkin labels.
This is the most explicit result for affine 
$su(4)$ fusion multiplicities that we know of.
The choice of order of summation in (\ref{sum4}) is immaterial, and we
recall the consistency conditions (\ref{S4})
which are required for a non-vanishing fusion multiplicity. They also ensure
that all bounds are integer.

\subsection{Threshold levels}

An obvious property of the convex polytope characterising the $su(4)$
fusion multiplicity is that it is connected. 
Furthermore, we have seen that the fusion polytope (\ref{cp4})
corresponds to ``slicing out'' a convex polytope embedded in the tensor
product polytope (\ref{ten4}). The slicing procedure involved then
shows us that the support of the threshold multiplicities
is also connected, i.e., for non-vanishing tensor product multiplicity
we have:
\ben
 {\rm For}\ \ T_{\la,\mu,\nu}>0:\ \ \ \ \ \ 
   n_{\la,\mu,\nu}^{(t)}>0\ \ \Longleftrightarrow\ \ t^\min\leq t\leq t^\max
\label{n}
\een
Even though this is a very natural result, we believe that our convex
polytope description has provided the most convincing evidence
hitherto. It is still not a rigorous proof since a rigorous
proof of (\ref{k04}) has yet to be found.

The argument leading to (\ref{n}) relies solely on the connectedness
of the tensor product polytope, and the assignment of a threshold level
to a true BZ triangle as in (\ref{k04}): $t$ is the {\em maximum}
of a set of expressions in the entries. In Ref. \cite{RW1}
we have shown that the tensor product multiplicities for all $su(r+1)$
may be characterised by convex polytopes, i.e., in particular
connected polytopes. Since we believe that
it is possible for all $su(r+1)$ to assign a 
threshold level to any true BZ triangle as a maximum of a set of expressions
in the entries, we conjecture that (\ref{n}) is valid for all $su(r+1)$.

The slicing procedure invites us to make a further conjecture concerning
the threshold multiplicities for a non-vanishing tensor product multiplicity:
\ben
 {\rm For}\ \ T_{\la,\mu,\nu}>0,\ \ \exists\  t_0\in\Z_\geq:\ \ \
 \left\{\begin{array}{ll}
   n_{\la,\mu,\nu}^{(t-1)}\leq n_{\la,\mu,\nu}^{(t)},\ \ \ t\leq t_0\\ \\ 
          n_{\la,\mu,\nu}^{(t)}\geq n_{\la,\mu,\nu}^{(t+1)},\ \ \ t>t_0
  \end{array}\right.
\label{t0}
\een
where $n_{\la,\mu,\nu}^{(t<0)}\equiv0$. This conjecture indicates 
that the threshold multiplicity $n_{\la,\mu,\nu}^{(t)}$ as a function 
of $t$ has exactly one local maximum. It is a refinement of (\ref{n})
since it implies (\ref{n}). It is made plausible by the observation that
the slicing procedure cuts off pieces of the polytope ``from above'':
all the affine conditions in (\ref{cp4}) correspond to planes with 
oriented normal vectors $(-1,0,0),\ (0,-1,0),\ (0,0,-1),\ (-1,-1,0),\ 
(-1,0,-1),\ (0,-1,-1)$ or $(-1,-1,-1)$. Since the polytope is {\em convex},
the slicing procedure for decreasing $k$ will in general
cut off bigger and bigger pieces until a certain point, after which the
pieces will become smaller and smaller. A maximum sized piece 
is cut off when $k$ decreases from $t_0$ to $t_0-1$. 
Note that $t_0$ may be any integer in the interval $[t^\min,t^\max]$. In 
general there will be a (non-vanishing and connected) 
sub-interval $[t^\min_0,t^\max_0]$, where 
any integer in it may play the role of $t_0$. In most of the cases we have
analysed explicitly, $t^\min_0=t^\max_0$, though that is not always true.
Examples are provided below.

We conjecture that (\ref{t0}) is valid for all $su(r+1)$. So far, this
conjecture has passed all the non-trivial tests we have made, though a
proof is still lacking.

The minimum threshold level $t^\min$ may be computed by determining 
necessary and sufficient conditions for 
the fusion multiplicity $N_{\la,\mu,\nu}^{(k)}$ to be non-vanishing.
Determining those, involves a straightforward, though cumbersome, 
analysis of the convex 
polytope (\ref{cp4}) or equivalently of the multiple sum formula (\ref{sum4})
-- see also \cite{RW1}. The resulting conditions may be expressed as a
set of inequalities in the (ordinary and dual) Dynkin labels and
the level $k$. We choose to present the result in terms of 
the Weyl group, since we believe that a similar and universal 
characterisation exists, valid for all simple Lie algebras. 

The Weyl group $W$ is generated by three simple reflections:
\ben
 s_i\la=\la-\la_i\al_i,\ \ i=1,2,3\ .
\label{Weyl}
\een
This extends readily to all simple Lie algebras.

Now, we find that $N_{\la,\mu,\nu}^{(k)}$ is non-vanishing provided
\bea
 0&\leq&\la_i,\ \mu_i,\ \nu_i,\ \ \ \ i=1,2,3,\nn
 0&\leq&\La^1\cdot(u\la+v\mu+w\nu),\ \ \ \ u,v,w\in
    \{I,\ s_1,\ s_1s_2,\ s_1s_2s_3\},\nn
  &&\ \ \ \ \ l(u)+l(v)+l(w)=3,\nn
 0&\leq&\La^2\cdot(u\la+v\mu+w\nu),\ \ \ \ u,v,w\in
    \{I,\ s_2,\ s_2s_1,\ s_2s_1s_3,\ s_2s_1s_3s_2\},\nn
  &&\ \ \ \ \ l(u)+l(v)+l(w)=4,\nn
 0&\leq&\La^2\cdot(u\la+v\mu+w\nu),\ \ \ \ u,v,w\in
    \{I,\ s_2,\ s_2s_3,\ s_2s_3s_1,\ s_2s_3s_1s_2\},\nn
  &&\ \ \ \ \ l(u)+l(v)+l(w)=4,\nn
 0&\leq&\La^3\cdot(u\la+v\mu+w\nu),\ \ \ \ u,v,w\in
    \{I,\ s_3,\ s_3s_2,\ s_3s_2s_1\},\nn
  &&\ \ \ \ \ l(u)+l(v)+l(w)=3
\label{nonv4}
\eea
and
\bea
 0&\leq&\la_0,\ \mu_0,\ \nu_0,\nn
 2k&\geq&S_2,\nn
 k&\geq& \La^1\cdot(u\la+v\mu+w\nu),\ \ \ \ u,v,w\in
    \{I,\ s_1,\ s_1s_2\},\ l(u)+l(v)+l(w)=2,\nn
 k&\geq&\La^2\cdot(u\la+v\mu+w\nu),\nn
 && \ (u,v,w)\in
    \{p(I,s_2,s_2s_1s_3),\ p(I,s_2s_1,s_2s_3),\ p(s_2,s_2,s_2s_1),\ 
   p(s_2,s_2,s_2s_3)\ |\ p\in {\cal S}_3\},\nn
 k&\geq&\La^3\cdot(u\la+v\mu+w\nu),\ \ \ \ u,v,w\in
    \{I,\ s_3,\ s_3s_2\},\ l(u)+l(v)+l(w)=2\ .
\label{nonv4k}
\eea
$I$ is the identity (i.e., $I\la=\la$), ${\cal S}_3$ 
is the permutation group of three elements, 
while $l(u)$ denotes the length of the
Weyl group element $u$ (with $l(I)=0$).
Note that the two $\La^2$-conditions in (\ref{nonv4}) contain identical
conditions on the weights. They are included to keep the expression
compact and symmetric. (\ref{nonv4}) contains altogether
50 independent constraints (in \cite{RW1} they are expressed as bounds on the
weight $\nu$), while (\ref{nonv4k}) contains 34 independent
constraints. 

To be clear, we stress that it is the Weyl orbits
of the fundamental weights which are important here. The 
Weyl group elements used above, and their particular expressions as
products of simple reflections $s_j$ (words), are not unique. 

At least some of the inequalities (\ref{nonv4}) and (\ref{nonv4k})
are quite easily understood. For example, 
$N^{(k)}_{\lambda,\mu,\nu}\not=0$ implies that $\nu^+-\lambda\in P(\mu)$, the 
set of weights (of non-vanishing multiplicity) of the module $M_\mu$. The 
boundaries of the weight diagram of $M_\mu$ are easily described:  
\ben
 \tilde\mu\in P(\mu)\ \Rightarrow\ 
  \mu\cdot\Lambda^j - \tilde\mu\cdot(w\Lambda^j)\ \ge\ 0\ ,\ \ 
  j=1,\ldots,r,\ w\in W\ .
\label{PmuW}
\een
Using $\nu^+=-w_0\nu$, we find 
\ben
 0\ \leq\ \Lambda^j\cdot\left(\, w\lambda +\mu +ww_0\nu \,\right)\ ,
\label{Wbdry}
\een 
for any $w\in W$. 
Other necessary inequalities can be written by permuting 
$\lambda,\mu,\nu$ in (\ref{Wbdry}). 

These are far from sufficient as conditions for non-vanishing tensor product 
multiplicities, however. In addition, they do not include any level-dependent 
constraints for fusion, like (\ref{nonv4k}). But fusion multiplicities 
may be found from certain formulas for tensor product multiplicities  
by replacing the 
Weyl group $W$ by the 
projection of the affine Weyl group. 
We suspect that the 
$k$-dependent inequalities may be explained this way. 

There is a long history 
of the problem of finding conditions on highest 
weights such that the corresponding tensor product multiplicity is non-zero. 
For a review, see \cite{Ful}. Recent advances include work on the 
tensor products for $GL_n(\C)$. Necessary and sufficient ``non-vanishing 
conditions'' turn out to have a relatively simple description in terms of 
Schubert calculus. The corresponding fusion problem has a similar 
solution, involving quantum Schubert calculus \cite{AW}. The connection 
between fusion and quantum cohomology goes back to Witten \cite{Wit}. 

As a simple application of (\ref{nonv4k}), $t^\min$ is found to be
\bea
 &&t^\min=\max\{\la^1+\la^3,\ \mu^1+\mu^3,\ \nu^1+\nu^3,\ [\hf(\la^2+\mu^2+
   \nu^2+1)],\nn
  &&\hspace{2cm}\la^1+\mu^1+\nu_3-\nu^3,\ \la^1+\mu_3-\mu^3+\nu^1,\ 
   \la_3-\la^3+\mu^1+\nu^1,\nn
  &&\hspace{2cm}\la^3+\mu^3+\nu_1-\nu^1,\ \la^3+\mu_1-\mu^1+\nu^3,\ 
   \la_1-\la^1+\mu^3+\nu^3,\nn
  &&\hspace{2cm}\la^1-\mu_1+\mu^1-\nu_1+\nu^1,\ -\la_1+\la^1+\mu^1-\nu_1
     +\nu^1,\ -\la_1+\la^1-\mu_1+\mu^1+\nu^1,\nn
  &&\hspace{2cm}\la^3+\mu^3-\mu_3+\nu^3-\nu_3,\ \la^3-\la_3+\mu^3+\nu^3
    -\nu_3,\ \la^3-\la_3+\mu^3-\mu_3+\nu^3,\nn
  &&\hspace{2cm}\la^2-\mu^1+\mu^3+\nu^1-\nu^3,\ -\la^1+\la^3+\mu^2+\nu^1
   -\nu^3,\ -\la^1+\la^3+\mu^1-\mu^3+\nu^2,\nn
  &&\hspace{2cm}\la^2+\mu^1-\mu^3-\nu^1+\nu^3,\ \la^1-\la^3+\mu^2-\nu^1
   +\nu^3,\ \la^1-\la^3-\mu^1+\mu^3+\nu^2,\nn
  &&\hspace{2cm}\la^2-\la_2-\mu^2+\mu_2+\nu^2,\ 
   -\la^2+\la_2+\mu^2-\mu_2+\nu^2,\ \la^2-\la_2+\mu^2-\nu^2+\nu_2,\nn
  &&\hspace{2cm}-\la^2+\la_2+\mu^2+\nu^2-\nu_2,\ 
   \la^2+\mu^2-\mu_2-\nu^2+\nu_2,\ \la^2-\mu^2+\mu_2+\nu^2-\nu_2,\nn
  &&\hspace{2cm}\la^2-\la_2+\mu^2-\mu_2-\nu^1+\nu^3,\ 
   \la^2-\la_2+\mu^2-\mu_2+\nu^1-\nu^3,\nn
  &&\hspace{2cm}\la^2-\la_2-\mu^1+\mu^3+\nu^2-\nu_2,\
   \la^2-\la_2+\mu^1-\mu^3+\nu^2-\nu_2,\nn 
  &&\hspace{2cm}-\la^1+\la^3+\mu^2-\mu_2+\nu^2-\nu_2,\
   \la^1-\la^3+\mu^2-\mu_2+\nu^2-\nu_2\}\ .
\label{tmin}
\eea

We find that the maximum threshold level $t^\max$ is
\ben
 t^\max=\min\{S_1,\ S_2-\max\{\la_2,\ \mu_2,\ \nu_2\},\ S_3\}
\label{tmax}
\een
This formula first appeared in \cite{Beg}, though without proof.
According to the general discussion below on upper bounds on $t^\max$,
(\ref{tmax}) states that the relevant bound (\ref{tupb})
for $su(4)$ is saturated. To prove that explicitly, 
we show that for non-vanishing tensor product multiplicity
$T_{\la,\mu,\nu}$, there is at least one integer point in the associated
convex polytope that corresponds to a (true) BZ triangle of threshold
level $t$ (\ref{k04}) equal to the conjectured value (\ref{tmax}).

Following the discussion above we should consider a point on
the surface of the polytope. The surface is characterised by
one or several of the BZ entries (\ref{trifour}) being zero.
When at least two entries vanish simultaneously, we are at
an intersection of faces. Since the vanishing of
a corner point ($m_{14}=0,\ n_{14}=0$
or $l_{14}=0$) corresponds to a minimum value of $v_1$, $v_2$ or $v_3$,
symmetry then allows us to consider three initial conditions only:
$n_{12}=0$, $m_{24}=0$ or $l_{23}=0$. To indicate how the proof
goes, let us assume that $n_{12}=0$. We may then add the sum 
${\cal V}_2+{\cal V}_3$ to the triangle (leaving $n_{12}=0$ unchanged)
until one of the entries
$n_{13}$, $m_{34}$, $l_{12}$, $n_{34}$, $m_{12}$, $l_{24}$ or
$m_{23}$ also vanish, or $m_{23}=1$. Depending on the situation one may then 
add additional numbers of ${\cal V}_2$ or ${\cal V}_3$ to obtain
yet another vanishing entry.
With sufficiently many vanishing entries, the threshold level
(\ref{k04}) always coincides with (\ref{tmax}). A priori, there
are many possible intersections to analyze, but a straightforward
and systematic approach makes the analysis tractable. In that way
we have shown that the maximum threshold level is given by the
simple expression (\ref{tmax}).

\subsection{Examples}

The threshold level is in general not linear in the BZ triangles:
\ben
 t({\cal T}_1+{\cal T}_2)\leq t({\cal T}_1)+t({\cal T}_2)
\label{t1t2}
\een
Here ${\cal T}_1$ and ${\cal T}_2$ indicate true BZ triangles.
Of particular interest is the behavior under addition of virtual
triangles. As is easily seen, the inequality (\ref{t1t2}) is in general not
saturated even in those cases:
\bea
 t({\cal T}+{\cal V}_i)&=&t({\cal T})+c,\ \ \ \ \ c\in\{0,\ 1\},\nn
 t({\cal T}+{\cal W}-{\cal V}_i)&=&t({\cal T})+c,\ \ \ \ \ 
    c\in\{0,\ 1,\ 2\},\nn
 t({\cal T}+{\cal W})&=&t({\cal T})+c,\ \ \ \ \ c\in\{1,\ 2\}\ ,
\label{tv}
\eea
whereas formally we may assign the threshold levels
\ben
 t({\cal V}_i)=1,\ \ \ \ \ \ t({\cal W}-{\cal V}_i)=t({\cal W})=2\ .
\label{tv2}
\een
Here we have introduced the linear combination
\ben
 {\cal W}\ \equiv\ {\cal V}_1+{\cal V}_2+{\cal V}_3\ =\ \matrix{1\cr
	\bar1~~\quad\bar1\cr
 0~\quad\quad ~~0\cr
 0~\quad\bar1\quad\bar1\quad ~0\cr
 \bar1\qquad\quad \bar1\qquad\quad \bar1 \cr
 1~\quad \bar1~\quad 0~\quad 0~\quad \bar1~\quad 1 \cr}
\label{W}
\een
Note that for $su(3)$ we have $t({\cal T}+{\cal V})=t({\cal T})+1$ for all
triangles. In the following we will discuss two examples where a
similarly simple situation occurs for $su(4)$.

Our first example is defined by $\la_2=\mu_2=0$, with the remaining 
seven non-negative Dynkin labels restricted a priori only by (\ref{S4}).
In this case the convex polytope is one-dimensional and generated by
${\cal W}$ (\ref{W}). It then follows from (\ref{n}) that
$t({\cal T}+{\cal W})=t({\cal T})+1$, and we conclude that we are in a 
situation similar to the general $su(3)$ case -- see (\ref{3bmw}) 
and (\ref{string}):
\ben
 N_{\la,\mu,\nu}^{(k<t^\max)}
  =\sum_{j=t^\min}^k1,\ \ \ \ \ \ \ 
   N_{\la,\mu,\nu}^{(k\geq t^\max)}=T_{\la,\mu,\nu}
  =\sum_{j=t^\min}^{t^\max}1\ .
\label{NT}
\een
The set of associated threshold levels is
\ben
 \{t^\min,\ t^\min+1,...,\ t^\max\}
\label{tt}
\een
In particular, all non-vanishing threshold multiplicities are 1, and
$t_0^\min=t^\min$ and $t_0^\max=t^\max$. Furthermore, it is easily shown that
for $\la_2=\mu_2=0$ the maximum threshold level (\ref{tmax}) is
\ben
 t^\max=S_2-\nu_2\ .
\een
It follows that the tensor product multiplicity may be expressed as
the minimum of a simple set of linear combinations of Dynkin labels:
\bea
 T_{\la,\mu,\nu}&=&1+\min\{\la_1,\ \la_3,\ \mu_1,\ \mu_3,\ \nu_1,\ \nu_3,\
   \la_1+\la_3-\nu_2,\ \mu_1+\mu_3-\nu_2,\nn
 &&\hspace{1cm}\la_1+\mu_3-\nu_2,\ \la_3+\mu_1-\nu_2,\ \la^2+\mu^2-\nu^2,\
  [\hf S_2]-\nu_2\nn
 &&\hspace{1cm}\la^2-\mu^2+\nu^2-\nu_2,\ -\la^2+\mu^2+\nu^2-\nu_2\nn
 &&\hspace{1cm}\la^1-\mu^1+\mu^2+\nu^1-\nu_2,\
  -\la^1+\la^2+\mu^1+\nu^1-\nu_2,\nn
 &&\hspace{1cm}\la^3+\mu^2-\mu^3+\nu^3-\nu_2,\
  \la^2-\la^3+\mu^3+\nu^3-\nu_2,\nn
 &&\hspace{1cm}-\la^1+\la^2-\mu^1+\mu^2+\nu^1,\ 
   \la^2-\la^3+\mu^2-\mu^3+\nu^3,\nn
 &&\hspace{1cm}\la^1-\mu^1+\mu^2-\nu^2+\nu^3,\ 
  -\la^1+\la^2+\mu^1-\nu^2+\nu^3,\nn
 &&\hspace{1cm}\la^3+\mu^2-\mu^3+\nu^1-\nu^2,\ 
  \la^2-\la^3+\mu^3+\nu^1-\nu^2
  \}\ , 
\label{Tex1}
\eea
using (\ref{tmin}). It is understood that $T_{\la,\mu,\nu}$ vanishes
if the expression (\ref{Tex1}) is negative.

Our second example is defined by $\la_3=\mu_3=0$, with the remaining 
seven non-negative Dynkin labels restricted a priori only by (\ref{S4}).
The convex polytope is one-dimensional and generated by
${\cal V}_1$ of (\ref{Vfour}). In the general expression (\ref{tmin})
for $t^\min$, one of the contributors is $\la^3-\la_3+\mu^3-\mu_3+\nu^3$
which in this case reduces to $S_3$. We may therefore conclude that 
\ben
 t^\min=t^\max=S_3
\een
It then follows that $t({\cal T}+{\cal V}_1)=t({\cal T})$, and there is
only one non-vanishing threshold multiplicity: 
$n^{(S_3)}_{\la,\mu,\nu}=T_{\la,\mu,\nu}$ and $t_0^\min=t_0^\max=S_3$.
A straightforward analysis of the single-sum expressing the tensor product
multiplicity yields 
\bea
 &&T_{\la,\mu,\nu}=1+\min\{\la_1,\ \mu_1,\ \nu_1,\ \nu_2,\ \nu_3,\ 
  -\la^1+\la^2-\mu^3+\nu^1,\ -\la^3-\mu^1+\mu^2+\nu^1,\nn
 &&\hspace{2cm}-\la^1+\mu^3+\nu^3,\ \la^3-\mu^1+\nu^3,\ 
   \la^1-\mu^3-\nu^1+\nu^2,\ -\la^3+\mu^1-\nu^1+\nu^2,\nn
 &&\hspace{2cm}\la^1-\la^3+\mu^1-\mu^3+\nu^1-\nu^3,\ 
   \la^1-\la^3+\mu^1-\mu^3-\nu^1+\nu^3,\nn
 &&\hspace{2cm}\la^1-\la^2+\mu^3+\nu^2-\nu^3,\ 
     \la^3+\mu^1-\mu^2+\nu^2-\nu^3,\ -\la^1+\la^3-\mu^1+\mu^3+\nu^2,\nn
 &&\hspace{2cm}\la^1-\la^3-\mu^1+\mu^3+\nu^1-\nu^2+\nu^3,\ 
   -\la^1+\la^3+\mu^1-\mu^3+\nu^1-\nu^2+\nu^3\}\ ,\nn  
\label{T2}
\eea 
provided $\la^3+\mu^3-\nu^1\geq0$ and the expression (\ref{T2})
is non-negative. If one of these conditions fails to be satisfied,
$T_{\la,\mu,\nu}$ vanishes by construction.

Furthermore, we can write a very simple but 
necessary condition on the tensor product 
multiplicity. Assuming the two conditions $\la^3+\mu^3-\nu^1\geq0$ and
$T_{\la,\mu,\nu}\geq0$ are satisfied, (\ref{T2}) easily yields 
\ben
 T_{\la,\mu,\nu}\leq1+\min\{\la_1,\ \la_2,\ \mu_1,\ \mu_2,\ \nu_1,\ \nu_2,\ 
   \nu_3\}\ .
\een

The two examples above illustrate two extreme situations: a ``horizontal''
distribution of threshold levels, and a ``vertical'' distribution.
In the first example we have a threshold level string with 
associated multiplicities all being one -- a purely horizontal distribution. 
In the second example we have exactly one non-vanishing threshold
multiplicity -- a purely vertical distribution. A generic situation
will have a much more complicated distribution, which we 
believe must respect (\ref{n}) and (\ref{t0}).

It is trivial to devise an alternative example of a purely vertical
distribution, namely $\la_1=\mu_1=0$. Analogues to the results above in the
case $\la_3=\mu_3=0$, are obtained by conjugation:
$(\la_1,\la_2,\la_3)^+=(\la_3,\la_2,\la_1)$, etc. Simple permutations of the
three weights $\la,\ \mu$ and $\nu$ also provide new examples.

\section{Refined depth rule and threshold levels} 

A refinement of the Gepner-Witten depth rule
\ben
 N_{\lambda,\mu,\nu}^{(k)}\ =\ \dim\{\, v\in M_{\mu,\nu^+-\lambda}\ |\ 
  f_i^{\ \nu_i+1}v=0,\ i=1,\ldots,r;\ e_\theta^{\ k-\nu\cdot\theta+1}v=0\,\}
\label{rdr}
\een 
was conjectured in \cite{KMSW,kmswii} 
(see also \cite{mwcjp}). Here $M_{\mu,\nu^+-\lambda}$ is the 
subspace of $M_\mu$ of weight $\nu^+-\lambda$, $f_i$ is the lowering operator 
associated to the simple root $\alpha_i$, and $e_\theta$ is the raising 
operator associated to $\theta$, the highest root of the simple Lie algebra 
(of rank $r$) being considered. When the level $k$ is large, the constraint 
$e_\theta^{\ k-\nu\cdot\theta+1}v=0$ is automatically satisfied, and 
a well-known formula for the tensor product coefficients is recovered. 
This agrees with (\ref{TNlim}).
  
In this section $u$, $v$ and $w$ denote vectors -- not Weyl group
elements as they did in the previous section.

We now derive an upper bound on the maximum threshold level associated to 
$N_{\lambda,\mu,\nu}^{(k)}\not=0$. The depth $d_v\in\Z_\ge$ of $v$ is the 
minimum possible power of $e_\theta$ such that $e_\theta^{\,d_v+1}v=0$.  
(\ref{rdr}) 
tells us that if a vector $v\in M_{\mu,\nu^+-\lambda}$ associated to 
a fusion coupling has depth $d_v$, then its threshold level is 
$t=d_v+\nu\cdot\theta$. 

An obvious upper 
bound on $d_v$ is the maximum number of times $\theta$ can be added to the 
weight $\nu^+-\lambda$ of $v$ to obtain a ``higher weight'' 
still having non-zero multiplicity in $M_\mu$. For the upper bound, 
the higher weight $\nu^+-\lambda+d_{v,{\rm max}}\theta$ should be a weight 
on the boundary of the weight diagram. Because $\theta$ is the highest root, 
the boundary weight will have 
$\Lambda^j\cdot(\nu^+-\lambda+d_{v,{\rm max}}\theta) 
= \Lambda^j\cdot\mu$, for some $j\in \{1,\ldots,r\}$. We write 
$\theta= \sum_{j=1}^r a^{\vee j}\alpha^\vee_j$, where $a^{\vee j}$ and 
$\alpha^\vee_j$ denote the $j$-th co-mark and simple co-root, respectively. 
Then we obtain  $d_{v,{\rm max}}=\min\{\, 
(\lambda^j+\mu^j-\nu^{j^+})/a^{\vee j}\ |\ j=1,\ldots,r\,\}\ ,$ 
where $\nu^{j^+}\equiv\Lambda^j\cdot\nu^+$. Finally, this gives 
\ben
 t^{\rm max}\ \leq\ 
  \min\{\, (\lambda^j+\mu^j-\nu^{j^+})/a^{\vee j}+\nu\cdot\theta\ |\ 
  j=1,\ldots,r\,\}\ .
\label{tup}
\een 
Note that all co-marks $a^\vee_j$ are non-vanishing.

This bound can be improved by noticing that (\ref{rdr}) treats the 
weights $\{\lambda,\mu,\nu\}$ asymmetrically. So permuting 
those weights will result in other upper bounds on $t^{\rm max}$. 
For $p\in{\cal S}_3$, 
let $p\{\lambda,\mu,\nu\}\equiv\{p\lambda,p\mu,p\nu\}$. Then we must have 
\ben
 t^{\rm max}\ \leq\ \min\, \{\, \left((p\lambda)^j+(p\mu)^j
    -(p\nu)^{j^+}\right)/a^{\vee j}+(p\nu)\cdot\theta\ |\ 
  j\in\{1,\ldots,r\},\ p\in{\cal S}_3\, \}\ . 
\label{tupb}
\een
Finally, we conjecture that this bound is saturated for all (untwisted)
affine Lie algebras, i.e., (\ref{tupb}) is an equality.
As shown in previous sections, that is true for $su(3)$ and $su(4)$,
and it is easily seen to be true for $su(2)$.

Also interesting is a 
similar argument based on the symmetric form of the depth rule. Suppose 
$u\in M_{\lambda,\bar\lambda}$, 
$v\in M_{\mu,\bar\mu}$,  
$w\in M_{\nu,\bar\nu}$, and that the Clebsch-Gordan coefficient 
$C_{u,v,w}\not=0$ 
for the coupling $M_\lambda\otimes M_\mu\otimes 
M_\nu \supset M_0$ of interest. Then $\bar\lambda+\bar\mu 
+\bar\nu = 0$, and the symmetric depth rule says that 
\ben
 (e_\theta)^au\otimes(e_\theta)^bv\otimes(e_\theta)^c
   w\ =\ 0\, ,\ \ {\rm for\ all}\ a+b+c>k\ .
\een
This implies that
\ben
 t^\max\ \leq\ \min\{\ (\la+\mu+\nu)^j/a^{\vee j}\ |\ j=1,\ldots,r\ \}\ . 
\label{tsym}
\een

The bound (\ref{tupb}) is always stronger than the bound (\ref{tsym}),
which is equivalent to stating that
\ben
 \nu\cdot\theta\ \leq\ (\nu^j+\nu^{j^+})/a^{\vee j}\ \ \ \ 
   {\rm for}\ j=1,\ldots,r\ .
\label{in}
\een
We have checked (\ref{in}) explicitly for all simple Lie algebras:
$A_r,\ B_r,\ldots,\ F_4,\ G_2$. 
Alternatively, one may regard the comparison of the bounds (\ref{tupb}) and
(\ref{tsym}) as an indirect proof of (\ref{in}), since from the point of 
view of the associated depth rules, the one leading to (\ref{tupb}) is
obviously the stronger one. We recall that $A_r\simeq su(r+1)$.

A weaker bound was previously found by Cummins for
$su(r+1)$. As quoted in \cite{BMW} it states that 
$t^\max\leq\min\{S_1,\ S_r\}$,
and was obtained using symmetric function techniques.
Here $S_1$ and $S_r$ are defined as in (\ref{S3}) and (\ref{S4}).

\section{Comments}

We anticipate that fusion multiplicities for higher rank $su(r+1)$ may
also be characterised by discretised polytope volumes. Whether or not such
a volume may be measured straightforwardly is less clear. One could 
imagine that complications such as the integer-value considerations above
would appear, and that they increase in complexity for higher rank.

We also believe that our approach may be extended to cover other simple
Lie algebras, and intend to discuss that elsewhere. Since a generalisation
of the BZ triangles to other Lie algebras is presently not known,
our belief is mainly based on our alternative approach to  
the computation of fusion multiplicities. 
It relies on the depth rule and the relation to three-point 
functions in Wess-Zumino-Witten 
conformal field theory \cite{Ras,RW4}. In that work
BZ triangles only appear as guidelines, while the basic building blocks
are certain polynomials. It is an
intriguing observation that the assignment of threshold levels to
those polynomials is straightforward, whereas the derivation of
(\ref{k04}) was quite cumbersome. 
One could therefore speculate that further progress, even for higher rank 
$su(r+1)$, is more likely to be made using the three-point function approach.

Along another 
line of generalisation, we are currently studying 
higher-point and higher-genus $su(3)$ and $su(4)$ fusions.
Thus, those efforts are combining the work presented here
with our recent results on higher-point and higher-genus $su(2)$ fusion
\cite{RW3}, and our general description of higher-point $su(r+1)$ tensor
product multiplicities \cite{RW2}.

Finally, when implemented in computer programs, we anticipate that our
multiple sum formulas offer a computational advantage over
more conventional methods such as Weyl group, and Young tableaux 
methods, for example.

\vskip.5cm
\noindent{\em Acknowledgements}
\vskip.1cm
\noindent We thank P. Mathieu for commenting on the manuscript, and
C. Woodward for informing us of reference  
\cite{AW}. The typing of this work was completed while J.R. visited the 
CRM, Montreal. He is grateful for its generous hospitality.

\end{document}